\documentclass[aps,prc,twocolumn,amsmath,amssymb,superscriptaddress,showpacs,showkeys]{revtex4}
\usepackage{dcolumn}
\usepackage{graphicx,color}
\usepackage{multirow}
\usepackage{textcomp}
\usepackage{physics}
\usepackage{relsize}
\usepackage{tikz}	
\usetikzlibrary
{
	arrows,
	calc,
	through,
	shapes.misc,	
	shapes.arrows,
	chains,
	matrix,
	intersections,
	positioning,	
	scopes,
	mindmap,
	shadings,
	shapes,
	backgrounds,
	decorations.text,
	decorations.markings,
	decorations.pathmorphing,	
}
\begin {document}
  \newcommand {\nc} {\newcommand}
  \nc {\beq} {\begin{eqnarray}}
  \nc {\eeq} {\nonumber \end{eqnarray}}
  \nc {\eeqn}[1] {\label {#1} \end{eqnarray}}
  \nc {\eol} {\nonumber \\}
  \nc {\eoln}[1] {\label {#1} \\}
  \nc {\ve} [1] {\mbox{\boldmath $#1$}}
  \nc {\ves} [1] {\mbox{\boldmath ${\scriptstyle #1}$}}
  \nc {\mrm} [1] {\mathrm{#1}}
  \nc {\half} {\mbox{$\frac{1}{2}$}}
  \nc {\thal} {\mbox{$\frac{3}{2}$}}
  \nc {\fial} {\mbox{$\frac{5}{2}$}}
  \nc {\la} {\mbox{$\langle$}}
  \nc {\ra} {\mbox{$\rangle$}}
  \nc {\etal} {\emph{et al.}}
  \nc {\eq} [1] {(\ref{#1})}
  \nc {\Eq} [1] {Eq.~(\ref{#1})}
  \nc {\Ref} [1] {Ref.~\cite{#1}}
  \nc {\Refc} [2] {Refs.~\cite[#1]{#2}}
  \nc {\Sec} [1] {Sec.~\ref{#1}}
  \nc {\chap} [1] {Chapter~\ref{#1}}
  \nc {\anx} [1] {Appendix~\ref{#1}}
  \nc {\tbl} [1] {Table~\ref{#1}}
  \nc {\fig} [1] {Fig.~\ref{#1}}
  \nc {\ex} [1] {$^{#1}$}
  \nc {\Sch} {Schr\"odinger }
  \nc {\flim} [2] {\mathop{\longrightarrow}\limits_{{#1}\rightarrow{#2}}}
  \nc {\textdegr}{$^{\circ}$}
  \nc {\inred} [1]{\textcolor{red}{#1}}
  \nc {\inblue} [1]{\textcolor{blue}{#1}}
  \nc {\IR} [1]{\textcolor{red}{#1}}
  \nc {\IB} [1]{\textcolor{blue}{#1}}
  \nc{\pderiv}[2]{\cfrac{\partial #1}{\partial #2}}
  \nc{\deriv}[2]{\cfrac{d#1}{d#2}}
\title{Low-energy corrections to the eikonal description of elastic scattering and breakup of one-neutron halo nuclei in nuclear-dominated reactions}
\author{C.~Hebborn}
\email{chloe.hebborn@ulb.ac.be}
\affiliation{Physique Nucl\' eaire et Physique Quantique (CP 229), Universit\'e libre de Bruxelles (ULB), B-1050 Brussels}
\author{P.~Capel}
\email{pcapel@uni-mainz.de}
\affiliation{Institut f\"ur Kernphysik, Johannes Gutenberg-Universit\"at Mainz, D-55099 Mainz}
\affiliation{Physique Nucl\' eaire et Physique Quantique (CP 229), Universit\'e libre de Bruxelles (ULB), B-1050 Brussels}
\date{\today}
\begin{abstract}
\begin{description}
\item[Background]
The eikonal approximation is a high-energy reaction model which is very computationally efficient and provides a simple interpretation of the collision.
Unfortunately, it is not valid at energies around 10~MeV/nucleon, the range of energy of HIE-ISOLDE  at CERN and the future ReA12 at MSU.
Fukui \etal\ [Phys. Rev. C {\bf 90}, 034617 (2014)] have shown that a simple semiclassical correction of the projectile-target deflection could improve the description of breakup of halo nuclei on heavy targets down to 20~MeV/nucleon.
\item[Purpose]
We study two similar corrections, which aim at improving the projectile-target relative motion within the eikonal approximation, with the goal to extend its range of validity down to 10~MeV/nucleon in nuclear-dominated collisions, viz. on light targets.
The semiclassical correction substitutes the impact parameter by the distance of closest approach of the corresponding classical trajectory.
The exact continued $S$-matrix correction replaces the eikonal phase  by the exact phase shift.
Both corrections successfully describe the elastic scattering of one-neutron halo nuclei.
\item[Method]
We extend these corrections and study their efficiency in describing the breakup channel.
We evaluate them in the case of $^{11}\mathrm{Be}$ impinging on $^{12}\mathrm{C}$ at 20 and 10~MeV/nucleon.
\item[Results]
Albeit efficient to reproduce the elastic channel, these corrections do not improve the description of the breakup of halo nuclei within the eikonal approximation down to 20~MeV/nucleon.
\item[Conclusions]
Our analysis of these corrections shows that improving the projectile-target relative motion is not the ultimate answer to extend the eikonal approximation down to low energies.
We suggest another avenue to reach this goal.

\end{description}
\end{abstract}
\maketitle
%


\section{Introduction}\label{Introduction}
During the last three decades, the development of Radioactive-Ion Beams (RIB) has favored the  discovery of exotic nuclei away from stability. Among these, halo nuclei exhibit the most peculiar structure~\cite{T96}. Due to their low-binding energy, one or two of their nucleons can tunnel far away from the core of the nucleus. These loosely bound nucleons hence form a diffuse halo around the core, leading to a matter radius much larger than stable nuclei~\cite{HJ87}.
Halo nuclei are therefore usually described within a few-body model: a core, which contains most of the nucleons, plus one or two valence nucleons.

Being located far from stability, they exhibit short lifetimes and cannot be studied through usual spectroscopic techniques.
Information about their structure is therefore usually inferred from reaction measurements.
To reliably analyze such experiments, an accurate reaction model coupled to a realistic description of the projectile is needed.
Several models have been developed to this aim~\cite{BC12}.  The Continuum Discretized Coupled Channel (CDCC) method can be considered as the state-of-the-art model~\cite{Kam86,TNT01,YOMM12}. It has the advantage to solve the few-body system within a fully quantal approach.
Its core idea is to expand the few-body wavefunction onto the eigenstates of the Hamiltonian describing the projectile internal structure, including both the bound states and the continuum, which, to be numerically tractable, needs to be discretized.
This leads to a set of coupled equations, whose resolution can be computationally challenging.

This motivates the use of approximations, such as the time-dependent approach and the eikonal model, that are less time-consuming than CDCC. The time-dependent resolution relies on a semiclassical description of the relative motion between the projectile and the target~\cite{BW81,AW75}, coupled with a quantum description of the projectile. This simplified approach can reproduce the breakup of one-neutron halo nuclei~\cite{KYS94,EBB95,TW99,Fal02,CBM03c} but misses some quantal effects in reaction observables~\cite{CEN12}.

Another computationally-affordable model is the eikonal approximation~\cite{G59}. It is built on the fact that at high energy the projectile-target wavefunction does not differ much from the incoming plane wave. This assumption leads to a simplification of the \Sch equation, which can therefore be solved more easily while keeping a quantum description of the collision. This model is mostly used to describe intermediate and high-energy collisions~\cite{HT03,BC12,ATT96,OYI03,B05,BCG05,OB10} but {in its usual form and without additional correction, it is not} suited for lower energies (below {40~MeV/nucleon}).

Some new facilities, such as HIE-ISOLDE at CERN or the upcoming ReA12 at MSU, are or will be able to provide RIBs at 10~MeV/nucleon. As CDCC exhibits convergence issues in this range of energy, extending the validity of the eikonal approximation to such energies would be of great interest. This  was  achieved for Coulomb-dominated reactions in Ref.~\cite{FOC14}, through the use of a semiclassical correction~\cite{BW81,AW75}. This semiclassical approach was generalized to the nuclear interaction for collisions involving two compact nuclei~\cite{AZV97}. We have shown in Ref.~\cite{HC17} that it also provides very precise results for the elastic scattering of one-neutron halo nuclei down to 10~MeV/nucleon.
In the present work, we investigate its efficiency to describe breakup observables on light targets with the hope to obtain in such a way a unified correction of the eikonal approximation for all types of targets.

The exact continued $S$-matrix correction~\cite{Wal73,BrookePhD,BAT99} is another way to improve the eikonal approximation. It was proven to be very precise for the description of elastic scattering of one-neutron halo nuclei~\cite{BAT99}. We study here its generalization to the breakup channel.

We present in Sec.~\ref{Sec2} the theoretical framework of the eikonal model and the two aforementioned corrections. Sec.~\ref{Sec3} focuses on the comparison of each correction with CDCC,   on the elastic scattering and breakup cross sections of $^{11}\mathrm{Be}$ impinging on a $^{12}\mathrm{C}$ at 20 and 10~MeV/nucleon. Our conclusions and prospects are summarized in Sec.~\ref{Conclusions}.

\section{Theoretical framework}\label{Sec2}

\subsection{Eikonal model}\label{Sec2A}
\begin{figure}
	\centering
	{\includegraphics[width=\linewidth]{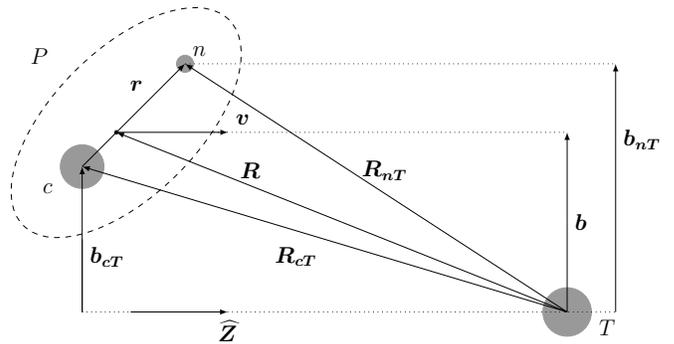}}
	\caption{\label{Fig3BodyCoordinates} Coordinates of the three-body model: the $c$-$n$ relative coordinate $\ve{r}$; the relative coordinate between the projectile center-of-mass and the target $\ve{R}$ and its component $\ve{b}$ transverse to the beam axis $\ve{\widehat Z}$, the $c$-$T$ and $n$-$T$ relative coordinates $\ve{R_{cT}}$ and $\ve{R_{nT}}$ with their transverse parts $\ve{b_{cT}}$ and $\ve{b_{nT}}$, respectively.}
\end{figure}

In this work, we focus on the elastic scattering and breakup of a one-neutron halo nucleus projectile $P$ impinging on a target $T$. 
As one-neutron halo nuclei exhibit a strong two-cluster structure, a reasonable description is to consider them as composed of a compact core $c$  to which a neutron $n$ is loosely bound~\cite{BC12}. This three-body model of the reaction is illustrated  in Fig.~\ref{Fig3BodyCoordinates}. For simplicity, we assume in the following all clusters to be spinless and structureless, and all the potentials to be central. 

The internal structure of the projectile is described by the effective $c$-$n$ Hamiltonian
\begin{equation}
h_{cn} = \frac{p^2 }{2 \mu_{cn}}  + V_{cn}(r).
\end{equation}
The kinetic  term $\frac{p^2 }{2 \mu_{cn}}$ depends on the $c$-$n$ relative momentum $\ve{p}$ and the $c$-$n$ reduced mass $\mu_{cn}$ while the $c$-$n$ interaction is simulated by a real effective potential $V_{cn}$, adjusted to reproduce the projectile low-energy spectrum. 

Studying this three-body collision hence corresponds to solving the following \Sch  equation
\begin{eqnarray}
\lefteqn{\left[\frac{P^2}{2\mu}+h_{cn}+V_{cT}(R_{cT})+V_{nT}(R_{nT})\right]\Psi(\ve{R},\ve{r})=}\nonumber\\
&\hspace{5.7cm}&E\ \Psi(\ve{R},\ve{r}), \label{eq2}
\end{eqnarray}
where $\ve{P}$ is the relative momentum between the projectile center-of-mass and the target and $\mu$ is the reduced mass of the whole three-body system.  As the projectile is modeled as a two-body object, two optical potentials are needed to simulate the interactions with the target. These potentials $V_{(c,n)T}$ depend only on the distance $R_{(c,n)T}$ between the projectile fragments $(c,n)$ and the target. The coordinates appearing in Eq.~\eqref{eq2} are defined in Fig.~\ref{Fig3BodyCoordinates}.

We consider that the projectile is initially in its ground state $\phi_0$, eigenstate of $h_{cn}$ of energy $\varepsilon_0$, and is impinging on the target with the wave vector $\ve{K}= K\ve{\hat{Z}}$, the $Z$-axis being chosen along the incoming beam. Therefore, Eq.~\eqref{eq2} has to be solved with the initial condition
\begin{equation}
\Psi(\ve{R},\ve{r})\flim{Z}{- \infty}\exp(iKZ+\cdots)\ \phi_0(\ve{r}),\label{eq3}
\end{equation}
where ``$\cdots$" reflects the fact that long-range interactions distort the incoming plane wave even at large distances.

At sufficiently high energy, the deviation from the incoming plane wave during the collision is small. In the eikonal model, this plane wave is factorized  out of the wavefunction~\cite{G59,BC12}
	\begin{equation}
	\Psi(\ve{R},\ve{r}) = \exp(i K Z)\ \widehat{\Psi}(\ve{R},\ve{r}).
	\end{equation}
Inserting this factorization in Eq.~\eqref{eq2} and assuming that the dependence in $\ve{R}$ of $\widehat\Psi$ is smooth enough to neglect its second-order derivatives compared to the first-order one, we obtain
	\begin{eqnarray}
	\lefteqn{i\hbar v \pderiv{}{Z} \widehat{\Psi}(\ve{b},Z,\ve{r})=}\nonumber\\
	&&\left[(h_{cn}-\varepsilon_0)+V_{cT}(R_{cT})+V_{nT}(R_{nT})\right]\widehat{\Psi}(\ve{b},Z,\ve{r}),\label{eq5}
	\end{eqnarray}
	where $v=\hbar K/\mu$ is the asymptotic $P$-$T$ relative velocity.
In \Eq{eq5}, the dependence of the three-body wavefunction $\widehat\Psi$ on $\ve{R}$ is explicitly decomposed into its longitudinal $Z$ and transverse $\ve{b}$ coordinates.
The description of the reaction through Eq.~\eqref{eq5} corresponds to the Dynamical Eikonal Approximation (DEA), which is very successful to model reactions involving one-nucleon halo nuclei at intermediate energy~\cite{BCG05,BC12}.
Note that by neglecting the second-order derivative along the transverse coordinate $\ve{b}$, part of the interference between neighboring $\ve{b}$ is missing in the eikonal approximation.
	
	The standard eikonal model makes an additional approximation, which considers the internal coordinates of the projectile $\ve{r}$ to be frozen during the collision. This assumption, called the adiabatic---or sudden---approximation, holds for collisions occurring in a very brief time, i.e. with only short-range interactions. In this adiabatic view $h_{cn}\approx \varepsilon_0$ and Eq.~\eqref{eq5} becomes
	\begin{equation}
	i\hbar v \pderiv{}{Z} \widehat{\Psi}^{\rm eik}(\ve{b},Z,\ve{r})=[V_{cT}(R_{cT})+V_{nT}(R_{nT})]\widehat{\Psi}^{\rm eik}(\ve{b},Z,\ve{r}).\label{eq6}
	\end{equation}
	The solutions of \eqref{eq6} satisfying the condition~\eqref{eq3}, read~\cite{G59,BC12}
	\begin{eqnarray}
	\lefteqn{\widehat{\Psi}^{\rm eik}(\ve{b},Z,\ve{r})=}\nonumber\\
	&&\exp \left[-\frac{i}{\hbar v} \int^Z_{-\infty} V_{cT} (\ve{b_{cT}},Z') +V_{nT} (\ve{b_{nT}},Z') 
	\mathrm{d}Z' \right] \phi_0(\ve{r}).\label{eq7}
\end{eqnarray}
 From a semiclassical viewpoint, these solutions describe the projectile following a straight-line trajectory at constant impact parameter $b$, along which the wavefunction accumulates a complex phase resulting from its interaction with the target. This phase can be split into two terms, one for each fragment of the projectile, which are computed for each  $\ve{b_{(c,n)T}}$ (see Fig.~\ref{Fig3BodyCoordinates}).
	
	After the collision, all the information about the reaction is carried in the eikonal phases
	\begin{equation}
	\chi_{(c,n)T}(\ve{b},\ve{r})=-\frac{1}{\hbar v}\int_{-\infty}^{+\infty} V_{(c,n)T}(\ve{R},\ve{r})\ \mathrm{d}Z.\label{eq8}
	\end{equation}
	The Coulomb interaction needs a particular treatment first because Eq.~\eqref{eq8} diverges for a Coulomb potential and second because its infinite range is incompatible with the adiabatic assumption. To solve these two issues, we use the Coulomb-corrected eikonal approximation (CCE)~\cite{MBB03,CBS08}.

Unfortunately, the simplicity and forward interpretation of the eikonal approximation work only at high enough energy, where the deviation from the initial plane wave is small.
At lower energy, we need to account for the deflection of the projectile due to its interaction with the target.
As shown in Ref.~\cite{FOC14}, this can be done for Coulomb-dominated collisions using a semiclassical correction.
We study here the low-energy extension of the eikonal approximation for nuclear-dominated reactions.

\subsection{Semiclassical correction}\label{Sec2B}
To account for the deflection of the projectile, we first study a semiclassical correction which replaces within the computation of the eikonal phase the impact parameter $b$ by the distance of closest approach $b'$ of the corresponding classical trajectory~\cite{BW81,AW75}. For a pure Coulomb collision, this distance $b'_C$  is known analytically~\cite{AW75,BW81}
	\begin{equation}
 b'_C = \frac{\eta + \sqrt{\eta^2 + \left(Kb\right)^2}}{K},
	\end{equation}
	where $\eta$ is the $P$-$T$ Sommerfeld parameter.
The fact that $b'_C>b$ accounts for the Coulomb repulsion between the projectile and the target. 
	Since this correction leads to very precise results for collisions with heavy targets~\cite{FOC14}, i.e. for Coulomb-dominated reactions, its extension to the nuclear interaction has been investigated \cite{HC17}.
	
	The nuclear deflection of the projectile can be taken into account by using the distance of closest approach $b'$ of the classical trajectory obtained with {both the Coulomb and} the real part of the nuclear optical potential{s} $V$ \cite{LVZ95,HCp17}
\begin{equation}
	E - \mathrm{Re}\left\{V(b')\right\} - \frac{\mu v^2 }{2}\left(\frac{b}{b'}\right)^2=0. \label{eq10}
	\end{equation}
	This implicit equation translates the conservation of energy and orbital angular momentum in the $P$-$T$ relative motion.
	We have observed that for nuclear-dominated collisions, this correction induces an increase of the eikonal elastic-scattering cross section that leads to a significant overestimation of the exact results. This suggests that the extension of the semiclassical correction to the real part of the nuclear interaction  misses absorption from the elastic channel~\cite{HCp17}.
	
	To tackle this issue, it has been proposed in Ref.~\cite{AZV97} to use a complex distance of closest approach $b''$ computed from the whole optical potential.
{This distance is solution of an equation similar to~\eqref{eq10}, but using {the complex} $V$ instead of its sole real part.
These $b''$ can be computed with numerical techniques, like Newton Raphson.
However, since the imaginary part of $b''$ remains small in all the cases studied here, its perturbative estimate obtained as the first iteration of the Newton-Raphson method is sufficient.
It reads~\cite{B85}}
	\begin{equation}
	b''=b'-i\,\left[\frac{\mathrm{Im}\left\{V(R)\right\}}{\deriv{}{R} \left(\mathrm{Re}\left\{V(R)\right\}+ E\frac{b^2}{R^2}\right)}\right]_{R=b'}, \label{eq11}
	\end{equation}
	where $b'$ is the solution of \Eq{eq10}.
	Including an imaginary part in the distance of closest approach enhances the absorption and hence improves the eikonal model.
	As it is very efficient  for elastic scattering of one- \cite{AZV97} and two-body projectiles \cite{HC17}, we study in the present paper its action on breakup observables.
	
	 At low energy, the adiabatic treatment of the collision becomes inadequate and the dynamical effects start to play a role~\cite{SAJ02}. 
To test the influence of the sudden approximation at the energies considered here, we extend the complex semiclassical correction to the DEA by replacing in \Eq{eq5} the transverse components of $\ve{R_{cT}}$ and $\ve{R_{nT}}$ by their corresponding complex distances of closest approach $b''_{cT}$ and $b''_{nT}$ obtained through~\Eq{eq11}.

\subsection{Exact continued $S$-matrix correction}\label{Sec2C}
Another way to correct the $P$-$T$ relative motion is to  use the exact correspondence between the partial-wave expansion and the eikonal model. Wallace demonstrates it  in the case of the elastic scattering of a structureless projectile~\cite{Wal73}.  He shows that the scattering amplitude of the partial-wave expansion can be computed by a series of integrals over the impact parameter $b$. The first term corresponds to the eikonal calculation wherein the eikonal phase is replaced by the exact phase shift
\begin{equation}
\chi(b) \to 2\delta_l ,
\end{equation}
relating the impact parameter to the orbital angular momentum through the semiclassical relation
\begin{equation}
l=Kb-1/2.\label{eq13}
\end{equation}

In Ref.~\cite{BAT99}, Brooke, Al-Khalili and Tostevin show that the zeroth-order of this sum is already very accurate to describe the elastic scattering of structureless nuclei.
They generalize this correction to the elastic scattering of one-neutron halo nuclei by assuming that, due to the spatial extension of the projectile {and the adiabatic treatment of their relative motion}, the total eikonal phase $\chi$ can be approached by the sum of the exact phase shifts of each fragment~\cite{BrookePhD}
\begin{equation}
\chi(\ve{R},\ve{r})=\chi_{cT}(\ve{b},\ve{r})+\chi_{nT}(\ve{b},\ve{r})\to 2\delta_{l_{cT}} +2\delta_{l_{nT} },\label{eq14}
\end{equation}
with $l_{cT}$ and $l_{nT}$, respectively, the $c$-$T$ and $n$-$T$ angular momenta related to the corresponding $b_{cT}$ and $b_{nT}$ of Fig.~\ref{Fig3BodyCoordinates} through Eq.~\eqref{eq13}.
{This correction hence substitutes the eikonal phase for each fragment of the projectile by its actual phase shift, leading to a more accurate description of their scattering by the target.}
It significantly enhances the precision of the eikonal description of elastic scattering down to 10~MeV/nucleons, accordingly we study in this paper its generalization to breakup reactions. 
{Following Brooke \etal\ \cite{BAT99}, we consider only the first term of Wallace's expansion and} replace the eikonal phases by the exact phase shifts within the computation of the breakup amplitude.
  
{In our calculations, we consider the phase shifts obtained at the closest integer $l$ to the value obtained through \Eq{eq13}.}
We have checked that using this rough interpolation leads to similar results as with phase shifts calculated for non-integer $l$ values.

\section{Results and discussion}\label{Sec3}
\subsection{Two-body interactions} \label{Sec3A}
To analyse  the corrections presented in Sec.~\ref{Sec2}, we perform calculations for the elastic scattering and breakup of $^{11}\mathrm{Be}$ impinging on $^{12}\mathrm{C}$ at 20 and 10~MeV/nucleon. The one-neutron halo nucleus $^{11}\mathrm{Be}$ is described as  an inert  $^{10}\mathrm{Be}$ core to which an $s$-valence neutron is bound by 0.5~MeV.
We simulate the $^{10}\mathrm{Be}$-${n}$ interaction by the Woods-Saxon potential given in Ref.~\cite{CGB04}, and adjust the depth to $V_R$=62.98~MeV to produce the 1/2+ ground state in the 1$s$ wave. We use the same potential in all partial waves but in the $d$ wave, where we use $V_R = 69.15$ MeV to account for the known 5/2+ resonance in the $^{10}\mathrm{Be}$-${n}$ continuum.
This potential produces a $d$ resonance at energy $E_{d}=1.27$~MeV and with a width $\Gamma_{d}=0.16$~MeV, which are close to the experimental values $E_{5/2^+}=1.274$~MeV and $\Gamma_{5/2^+}=100$~keV.
The $^{10}\mathrm{Be}$-$^{12}\mathrm{C}$ and ${n}$-$^{12}\mathrm{C}$ interactions are simulated by the same potentials  as in Ref.~\cite{HC17}\footnote{Note that in Ref.~\cite{HC17} the depth of the surface imaginary term of the ${n}$-$^{12}\mathrm{C}$ potential of Ref.~\cite{KD03} is erroneous and should read $W_D=7.1585$~MeV.}.

\subsection{Analysis}\label{Sec3B}
\begin{figure*}
	\center
	{\includegraphics[width=0.48\linewidth]{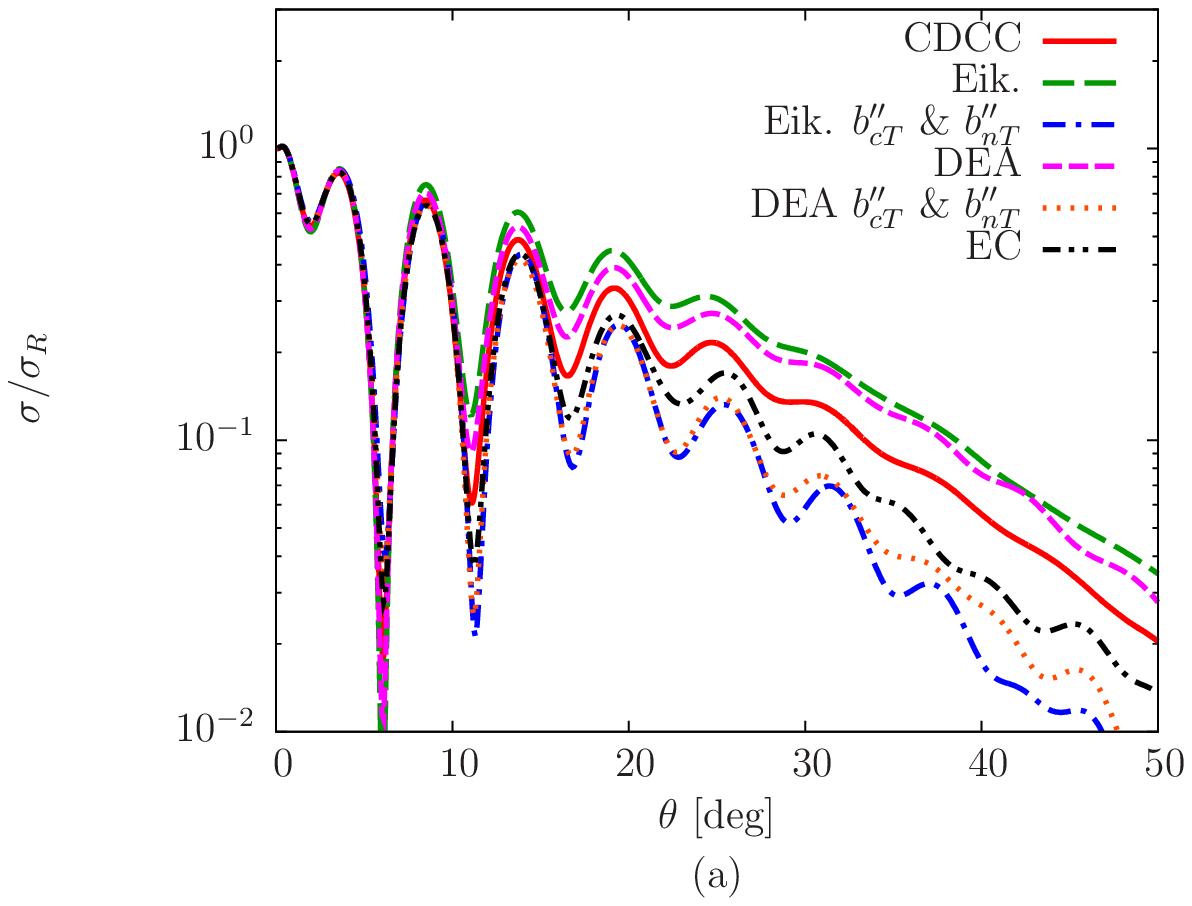}}
	\hspace{0.3cm}
	{	\includegraphics[width=0.48\linewidth]{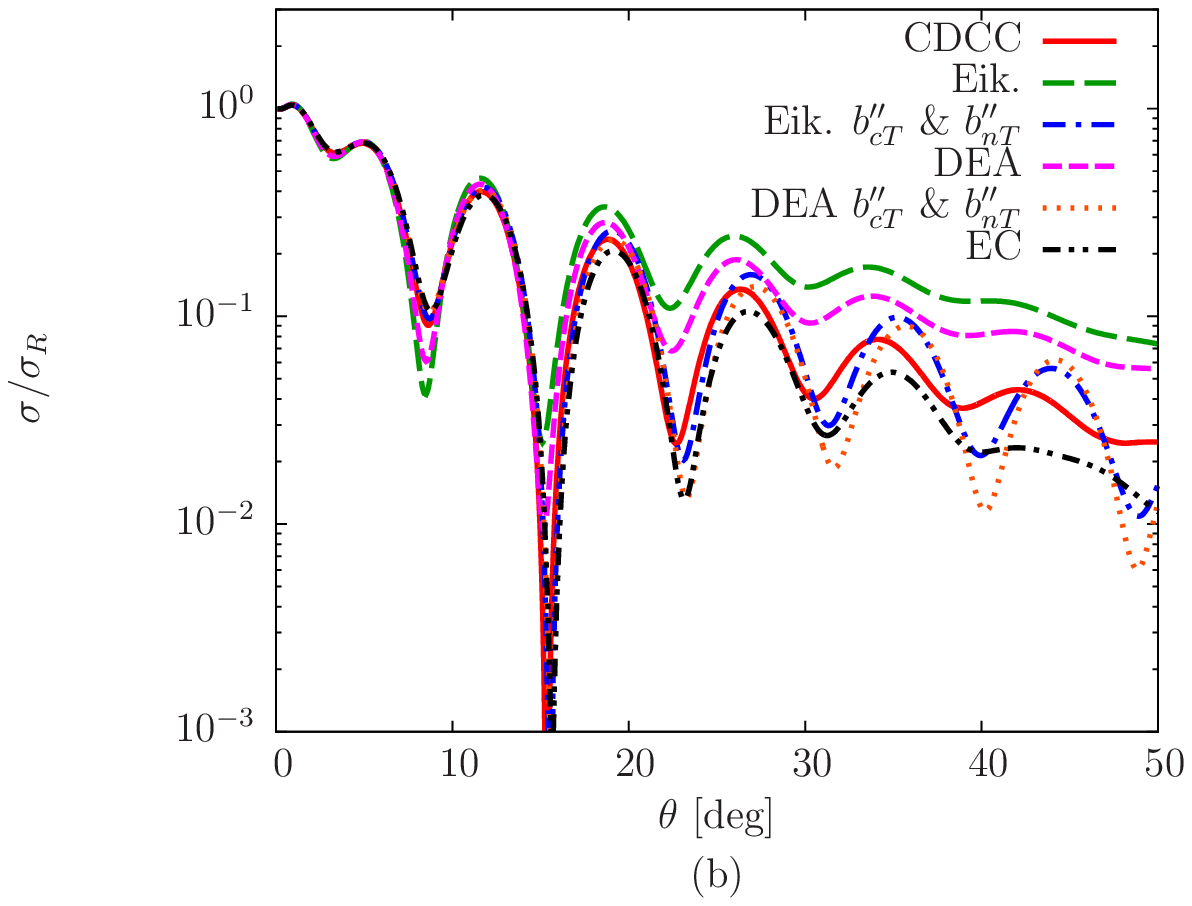}}
	\caption{Rutherford-normalized elastic-scattering cross sections of $^{11}\mathrm{Be}$ off $^{12}\mathrm{C}$ at  (a) 20~MeV/nucleon and (b) 10~MeV/nucleon as a function of the scattering angle $\theta$. The results are obtained with the CDCC, the standard eikonal approximation, the DEA and various corrections detailed in the text.
	}
	\label{Fig11Be2010AMeVSel}
\end{figure*}

{To estimate the quality of the corrections mentioned in the previous section, we confront their predictions to CDCC computations that we take as reference.
We perform these calculations with {\sc Fresco}~\cite{FRESCO} and use the following model space: the $^{10}\mathrm{Be}-n$ continuum is described up to the $c$-$n$ orbital angular momentum $l_{\rm max}=6$, the maximum $c$-$n$ energy is set to be $E_{\rm max}=10$~MeV, the number of bin states per partial waves is between $11$ (for
large $l$), 25 (at low $l$) and up to $49$ (within the $d$ wave to account for the presence of the resonance), and the total angular momentum is considered up to  $J_{\rm max}=20\,000$.} 

{ We plot in Fig.~\ref{Fig11Be2010AMeVSel} the differential elastic-scattering cross sections at (a)~20~MeV/nucleon and (b)~10~MeV/nucleon.}
At large angles, the eikonal cross sections (dashed green lines) overestimate the CDCC calculations {(solid red lines)}, suggesting that this model underestimates the absorption from the elastic channel.
It also tends to dampen the oscillations in that angular region.
Interestingly, the DEA cross sections (short dashed magenta lines) lie close to the eikonal ones, indicating that the dynamics of the projectile has little effect on the elastic-scattering process.
Note however that these dynamical effects increase at low energy since the discrepancy between the DEA and the eikonal results is larger at 10~MeV/nucleon than at 20~MeV/nucleon, as already observed in Ref.~\cite{SAJ02}.
 
Let us first compare the complex semiclassical correction~\eqref{eq11} within the standard eikonal model (dash-dotted blue lines) and  the DEA (dotted orange lines). 
In both cases, this correction leads to very similar cross sections.
At forward angles (below 15$^\circ$ and 20$^\circ$ at 20 and 10~MeV/nucleon, respectively), it improves the description of the elastic scattering.
At larger angles, it overcorrects the magnitude of the cross sections, which now falls below the reference calculation.
The oscillatory pattern of the CDCC calculation is no well reproduced.
Surprisingly, the magnitude of the cross sections obtained with this semiclassical correction are in better agreement with the CDCC results at 10~MeV/nucleon than at 20~MeV/nucleon.

Contrary to the semiclassical correction, the exact continued $S$-matrix one~\eqref{eq14} (dash-dotted-dotted black lines) has a similar accuracy at both energies. The magnitude of the cross sections is well reproduced up to 15$^\circ$ and 20$^\circ$ at   20 and 10~MeV/nucleon, respectively. At larger angles, they slightly underestimate the CDCC calculations. In addition, the oscillatory pattern is precise up to 25$^\circ$ at both energies, but is slightly shifted to larger angles. Our analysis indicates that this shift is due to the adiabatic assumption, still considered in this correction.  The remaining discrepancy with CDCC probably comes from the higher-order terms in the series of integrals over $b$ developed by Wallace, that we neglect in our computation (see Sec.~\ref{Sec2C}). 

These two corrections improve the accuracy of the eikonal description of the elastic scattering of halo nuclei at low energies, and, as such, properly correct the deflection of the projectile by the target.
In particular, they better simulate the absorption from the elastic channel and reproduce the oscillatory pattern of the reference calculation.
Moreover, these two corrections are insensitive to the choice of optical potentials. 
These encouraging results have driven us to extend them to breakup reactions.

\begin{figure}
	\center
	{\includegraphics[width=\linewidth]{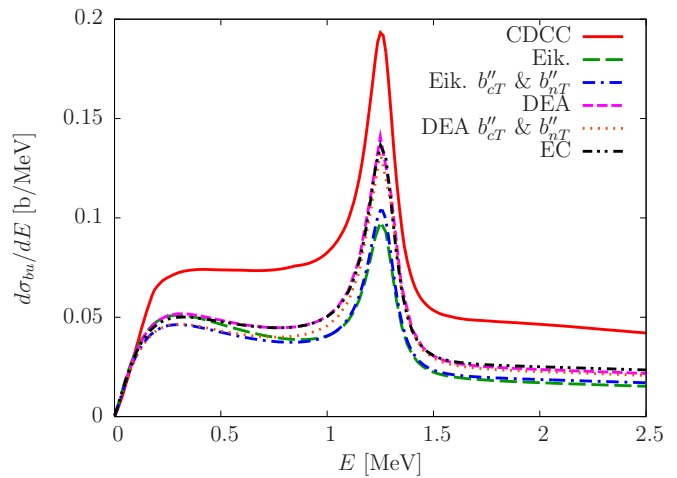}}
	\hspace{0.3cm}
	\caption{Breakup cross sections of $^{11}\mathrm{Be}$ impinging on $^{12}\mathrm{C}$ at 20~MeV/nucleon as a function  of the $^{10}\mathrm{Be}$-${n}$ relative energy.  }
	\label{Fig10Be20AMeVSbu}
\end{figure}

 In Fig.~\ref{Fig10Be20AMeVSbu}, we plot the breakup cross section of $^{11}\mathrm{Be}$ on $^{12}\mathrm{C}$ at  20~MeV/nucleon as a function of the $^{10}\mathrm{Be}$-${n}$ relative energy. Compared to CDCC, the eikonal model reproduces the right shape of the distribution, except in the range between 0.25~MeV and 1~MeV, where it predicts a local minimum. Unfortunately, it underestimates the cross sections by approximatively 30\%, over the whole energy range.  As in the elastic scattering case, including the dynamics does not impact significantly the accuracy: the DEA sure leads to a slightly larger cross section and improves the shape of the distribution between 0.25~MeV and 1~MeV, but this is far from enough to reach the CDCC prediction.

Figure~\ref{Fig10Be20AMeVSbu} shows that the semiclassical correction acts  similarly when applied to the standard eikonal approximation and to the DEA in breakup observables. It deteriorates the accuracy of these models, since the cross sections are reduced below 1~MeV and that there is no improvement at larger energy.
{This suggests that using the complex distance of closest approach increases the absorption from all the reaction channels and not just in the elastic-scattering one.
This correction should thus be treated with care as it worsens the eikonal description of breakup.}
Interestingly, the exact continued $S$-matrix correction gives results very close to the DEA, although it still relies on the adiabatic approximation.
This suggests that part of the projectile dynamics is restored through that correction.
{This approach has probably more potential as both reaction channels are improved simultaneously.}

Although these two corrections are very accurate to describe elastic scattering, they fail to reproduce the CDCC breakup cross sections.
{These negative results have been obtained at both 20 (see \fig{Fig10Be20AMeVSbu}) and 10~MeV/nucleon (not shown here).}
This indicates that improving the deflection of the projectile by the target is not the only issue that should be addressed to correct the eikonal approximation on light targets at low beam energy.
As mentioned in \Sec{Sec2A}, this approximation, neglecting the second-order derivatives in $\ve{b}$, misses part of the couplings between the angular momenta, i.e. between the  ``trajectories" at different impact parameters. {
Therefore, another way to improve the eikonal description of breakup at low energies {could} be to include the missing couplings between the angular momenta, e.g. perturbatively.}

\section{Conclusions}\label{Conclusions}
Being away from stability, halo nuclei can be studied only through indirect techniques, such as reactions. To interpret precisely the  measurements, one needs an accurate reaction model.  Nowadays, some laboratories are able to deliver RIBs at about 10~MeV/nucleon. In this range of energy, CDCC is very computationally-expensive and can present convergence issues. The eikonal model has the advantages to be much faster and to provide a simple interpretation of the collision. Unfortunately, it is  valid only at high energy.
In the case of Coulomb-dominated reactions, it has been extended down to low-energy thanks to a semiclassical correction \cite{FOC14}. More recently we have achieved the same for the elastic scattering on light targets, i.e. nuclear-dominated collisions \cite{HC17}. The present article focuses on the extension of these corrections to nuclear breakup at low energies.

The semiclassical correction, already proven successful for the elastic scattering of one-neutron halo nuclei~\cite{HC17}, aims to improve the deflection of the projectile by the target. It replaces the impact parameter by the  complex distance of closest approach of the corresponding classical trajectory~\cite{AZV97}. Although its generalization to breakup reactions is straightforward, it {tends to deteriorate} the accuracy of the eikonal model.
Interestingly, we have shown that for these nuclear-dominated reactions, the dynamics of the projectile has little effects on both elastic-scattering and breakup observables.

Another way to better simulate the deflection of the projectile by the target is to use the exact continued $S$-matrix correction~\cite{BrookePhD,Wal73,BAT99}. It consists in the first term of the exact correspondence between the partial-wave expansion and the eikonal model, derived by Wallace in the case of structureless nuclei~\cite{Wal73}. As for the semiclassical corrections, previous work~\cite{BAT99} has demonstrated its efficiency for the elastic scattering of one-neutron halo nuclei. In the present paper, we have generalized it to breakup reactions, but unfortunately this correction brings only a small accuracy gain in the description of nuclear breakup at the eikonal approximation.

These two corrections, focusing on the deflection of the projectile by the target,  extend the validity of the eikonal model only for the elastic scattering and not for breakup down to energies reachable by HIE-ISOLDE and ReA12. {We therefore believe that to improve the eikonal description of breakup observables, another flaw of the model should be tackled, which might {be} the underestimation of the couplings between the angular momenta, caused by the neglect of the second-order derivatives in the \Sch equation. Using a perturbative approach to account for these missed couplings could solve this issue.}


\begin{acknowledgements}
	We thank J. A. Tostevin for pointing to us the existence of the exact $S$-matrix correction and Ref.~\cite{BAT99}.
	C.~Hebborn acknowledges the support of the Fund for Research Training in Industry and Agriculture (FRIA), Belgium.   This project has received funding from the European Union’s Horizon 2020 research
	and innovation program under grant agreement
	No 654002, the Deutsche Forschungsgemeinschaft within the Collaborative
	Research Centers 1245 and 1044, and the
	PRISMA (Precision Physics, Fundamental Interactions and Structure of Matter) Cluster of Excellence. P. C. acknowledges the support of the State of Rhineland-Palatinate.
\end{acknowledgements}

\bibliographystyle{apsrev}
\bibliography{PRCCorrSCTost_BiblioR1}

\begin{thebibliography}{36}
\expandafter\ifx\csname natexlab\endcsname\relax\def\natexlab#1{#1}\fi
\expandafter\ifx\csname bibnamefont\endcsname\relax
  \def\bibnamefont#1{#1}\fi
\expandafter\ifx\csname bibfnamefont\endcsname\relax
  \def\bibfnamefont#1{#1}\fi
\expandafter\ifx\csname citenamefont\endcsname\relax
  \def\citenamefont#1{#1}\fi
\expandafter\ifx\csname url\endcsname\relax
  \def\url#1{\texttt{#1}}\fi
\expandafter\ifx\csname urlprefix\endcsname\relax\def\urlprefix{URL }\fi
\providecommand{\bibinfo}[2]{#2}
\providecommand{\eprint}[2][]{\url{#2}}

\bibitem[{\citenamefont{Tanihata}(1996)}]{T96}
\bibinfo{author}{\bibfnamefont{I.}~\bibnamefont{Tanihata}},
  \bibinfo{journal}{J.~Phys.~G} \textbf{\bibinfo{volume}{22}},
  \bibinfo{pages}{157} (\bibinfo{year}{1996}).

\bibitem[{\citenamefont{Hansen and Jonson}(1987)}]{HJ87}
\bibinfo{author}{\bibfnamefont{P.~G.} \bibnamefont{Hansen}} \bibnamefont{and}
  \bibinfo{author}{\bibfnamefont{B.}~\bibnamefont{Jonson}},
  \bibinfo{journal}{Europhys. Lett.} \textbf{\bibinfo{volume}{4}},
  \bibinfo{pages}{409} (\bibinfo{year}{1987}).

\bibitem[{\citenamefont{Baye and Capel}(2012)}]{BC12}
\bibinfo{author}{\bibfnamefont{D.}~\bibnamefont{Baye}} \bibnamefont{and}
  \bibinfo{author}{\bibfnamefont{P.}~\bibnamefont{Capel}},
  \emph{\bibinfo{title}{Breakup reaction models for two- and three-cluster
  projectiles}} (\bibinfo{publisher}{Springer}, \bibinfo{address}{Heidelberg},
  \bibinfo{year}{2012}), vol. \bibinfo{volume}{848} of
  \emph{\bibinfo{series}{Lecture Notes in Physics}}, \bibinfo{note}{pp.
  121--163}.

\bibitem[{\citenamefont{Kamimura et~al.}(1986)\citenamefont{Kamimura, Yahiro,
  Iseri, Kameyama, Sakuragi, and Kawai}}]{Kam86}
\bibinfo{author}{\bibfnamefont{M.}~\bibnamefont{Kamimura}},
  \bibinfo{author}{\bibfnamefont{M.}~\bibnamefont{Yahiro}},
  \bibinfo{author}{\bibfnamefont{Y.}~\bibnamefont{Iseri}},
  \bibinfo{author}{\bibfnamefont{H.}~\bibnamefont{Kameyama}},
  \bibinfo{author}{\bibfnamefont{Y.}~\bibnamefont{Sakuragi}}, \bibnamefont{and}
  \bibinfo{author}{\bibfnamefont{M.}~\bibnamefont{Kawai}},
  \bibinfo{journal}{Prog. Theor. Phys. Suppl.} \textbf{\bibinfo{volume}{89}},
  \bibinfo{pages}{1} (\bibinfo{year}{1986}).

\bibitem[{\citenamefont{Tostevin et~al.}(2001)\citenamefont{Tostevin, Nunes,
  and Thompson}}]{TNT01}
\bibinfo{author}{\bibfnamefont{J.~A.} \bibnamefont{Tostevin}},
  \bibinfo{author}{\bibfnamefont{F.~M.} \bibnamefont{Nunes}}, \bibnamefont{and}
  \bibinfo{author}{\bibfnamefont{I.~J.} \bibnamefont{Thompson}},
  \bibinfo{journal}{Phys. Rev. C} \textbf{\bibinfo{volume}{63}},
  \bibinfo{pages}{024617} (\bibinfo{year}{2001}).

\bibitem[{\citenamefont{Yahiro et~al.}(2012)\citenamefont{Yahiro, Ogata,
  Matsumoto, and Minomo}}]{YOMM12}
\bibinfo{author}{\bibfnamefont{M.}~\bibnamefont{Yahiro}},
  \bibinfo{author}{\bibfnamefont{K.}~\bibnamefont{Ogata}},
  \bibinfo{author}{\bibfnamefont{T.}~\bibnamefont{Matsumoto}},
  \bibnamefont{and} \bibinfo{author}{\bibfnamefont{K.}~\bibnamefont{Minomo}},
  \bibinfo{journal}{Prog. Theor. Exp. Phys.} \textbf{\bibinfo{volume}{2012}},
  \bibinfo{pages}{01A206} (\bibinfo{year}{2012}).

\bibitem[{\citenamefont{Broglia and Winther}(1981)}]{BW81}
\bibinfo{author}{\bibfnamefont{R.~A.} \bibnamefont{Broglia}} \bibnamefont{and}
  \bibinfo{author}{\bibfnamefont{A.}~\bibnamefont{Winther}},
  \emph{\bibinfo{title}{Heavy Ion Reactions, Lectures Notes, Vol. 1: Elastic
  and Inelastic Reactions}} (\bibinfo{publisher}{Benjamin-Cummings},
  \bibinfo{address}{Reading, England}, \bibinfo{year}{1981}).

\bibitem[{\citenamefont{Alder and Winther}(1975)}]{AW75}
\bibinfo{author}{\bibfnamefont{K.}~\bibnamefont{Alder}} \bibnamefont{and}
  \bibinfo{author}{\bibfnamefont{A.}~\bibnamefont{Winther}},
  \emph{\bibinfo{title}{Electromagnetic Excitation}}
  (\bibinfo{publisher}{North-Holland}, \bibinfo{address}{Amsterdam},
  \bibinfo{year}{1975}).

\bibitem[{\citenamefont{Kido et~al.}(1994)\citenamefont{Kido, Yabana, and
  Suzuki}}]{KYS94}
\bibinfo{author}{\bibfnamefont{T.}~\bibnamefont{Kido}},
  \bibinfo{author}{\bibfnamefont{K.}~\bibnamefont{Yabana}}, \bibnamefont{and}
  \bibinfo{author}{\bibfnamefont{Y.}~\bibnamefont{Suzuki}},
  \bibinfo{journal}{Phys. Rev. C} \textbf{\bibinfo{volume}{50}},
  \bibinfo{pages}{R1276} (\bibinfo{year}{1994}).

\bibitem[{\citenamefont{Esbensen et~al.}(1995)\citenamefont{Esbensen, Bertsch,
  and Bertulani}}]{EBB95}
\bibinfo{author}{\bibfnamefont{H.}~\bibnamefont{Esbensen}},
  \bibinfo{author}{\bibfnamefont{G.~F.} \bibnamefont{Bertsch}},
  \bibnamefont{and} \bibinfo{author}{\bibfnamefont{C.~A.}
  \bibnamefont{Bertulani}}, \bibinfo{journal}{Nucl. Phys. {\textbf{A}}}
  \textbf{\bibinfo{volume}{581}}, \bibinfo{pages}{107} (\bibinfo{year}{1995}).

\bibitem[{\citenamefont{Typel and Wolter}(1999)}]{TW99}
\bibinfo{author}{\bibfnamefont{S.}~\bibnamefont{Typel}} \bibnamefont{and}
  \bibinfo{author}{\bibfnamefont{H.~H.} \bibnamefont{Wolter}},
  \bibinfo{journal}{Z. Naturforsch. Teil A} \textbf{\bibinfo{volume}{54}},
  \bibinfo{pages}{63} (\bibinfo{year}{1999}).

\bibitem[{\citenamefont{Fallot et~al.}(2002)\citenamefont{Fallot, Scarpaci,
  Lacroix, Chomaz, and Margueron}}]{Fal02}
\bibinfo{author}{\bibfnamefont{M.}~\bibnamefont{Fallot}},
  \bibinfo{author}{\bibfnamefont{J.~A.} \bibnamefont{Scarpaci}},
  \bibinfo{author}{\bibfnamefont{D.}~\bibnamefont{Lacroix}},
  \bibinfo{author}{\bibfnamefont{P.}~\bibnamefont{Chomaz}}, \bibnamefont{and}
  \bibinfo{author}{\bibfnamefont{J.}~\bibnamefont{Margueron}},
  \bibinfo{journal}{Nucl. Phys. {\textbf{A}}} \textbf{\bibinfo{volume}{700}},
  \bibinfo{pages}{70} (\bibinfo{year}{2002}).

\bibitem[{\citenamefont{Capel et~al.}(2003)\citenamefont{Capel, Baye, and
  Melezhik}}]{CBM03c}
\bibinfo{author}{\bibfnamefont{P.}~\bibnamefont{Capel}},
  \bibinfo{author}{\bibfnamefont{D.}~\bibnamefont{Baye}}, \bibnamefont{and}
  \bibinfo{author}{\bibfnamefont{V.~S.} \bibnamefont{Melezhik}},
  \bibinfo{journal}{Phys. Rev. C} \textbf{\bibinfo{volume}{68}},
  \bibinfo{pages}{014612} (\bibinfo{year}{2003}).

\bibitem[{\citenamefont{Capel et~al.}(2012)\citenamefont{Capel, Esbensen, and
  Nunes}}]{CEN12}
\bibinfo{author}{\bibfnamefont{P.}~\bibnamefont{Capel}},
  \bibinfo{author}{\bibfnamefont{H.}~\bibnamefont{Esbensen}}, \bibnamefont{and}
  \bibinfo{author}{\bibfnamefont{F.~M.} \bibnamefont{Nunes}},
  \bibinfo{journal}{Phys. Rev. C} \textbf{\bibinfo{volume}{85}},
  \bibinfo{pages}{044604} (\bibinfo{year}{2012}).

\bibitem[{\citenamefont{Glauber}(1959)}]{G59}
\bibinfo{author}{\bibfnamefont{R.~J.} \bibnamefont{Glauber}}, in
  \emph{\bibinfo{booktitle}{Lecture in Theoretical Physics}}, edited by
  \bibinfo{editor}{\bibfnamefont{W.~E.} \bibnamefont{Brittin}}
  \bibnamefont{and} \bibinfo{editor}{\bibfnamefont{L.~G.} \bibnamefont{Dunham}}
  (\bibinfo{publisher}{Interscience}, \bibinfo{address}{New York},
  \bibinfo{year}{1959}), vol.~\bibinfo{volume}{1}, p. \bibinfo{pages}{315}.

\bibitem[{\citenamefont{Hansen and Tostevin}(2003)}]{HT03}
\bibinfo{author}{\bibfnamefont{P.~G.} \bibnamefont{Hansen}} \bibnamefont{and}
  \bibinfo{author}{\bibfnamefont{J.~A.} \bibnamefont{Tostevin}},
  \bibinfo{journal}{Ann. Rev. Nucl. Part. Sc.} \textbf{\bibinfo{volume}{53}},
  \bibinfo{pages}{219} (\bibinfo{year}{2003}).

\bibitem[{\citenamefont{Al-Khalili et~al.}(1996)\citenamefont{Al-Khalili,
  Tostevin, and Thompson}}]{ATT96}
\bibinfo{author}{\bibfnamefont{J.~S.} \bibnamefont{Al-Khalili}},
  \bibinfo{author}{\bibfnamefont{J.~A.} \bibnamefont{Tostevin}},
  \bibnamefont{and} \bibinfo{author}{\bibfnamefont{I.~J.}
  \bibnamefont{Thompson}}, \bibinfo{journal}{Phys. Rev. C}
  \textbf{\bibinfo{volume}{54}}, \bibinfo{pages}{1843} (\bibinfo{year}{1996}).

\bibitem[{\citenamefont{Ogata et~al.}(2003)\citenamefont{Ogata, Yahiro, Iseri,
  Matsumoto, and Kamimura}}]{OYI03}
\bibinfo{author}{\bibfnamefont{K.}~\bibnamefont{Ogata}},
  \bibinfo{author}{\bibfnamefont{M.}~\bibnamefont{Yahiro}},
  \bibinfo{author}{\bibfnamefont{Y.}~\bibnamefont{Iseri}},
  \bibinfo{author}{\bibfnamefont{T.}~\bibnamefont{Matsumoto}},
  \bibnamefont{and} \bibinfo{author}{\bibfnamefont{M.}~\bibnamefont{Kamimura}},
  \bibinfo{journal}{Phys. Rev. C} \textbf{\bibinfo{volume}{68}},
  \bibinfo{pages}{064609} (\bibinfo{year}{2003}).

\bibitem[{\citenamefont{Bertulani}(2005)}]{B05}
\bibinfo{author}{\bibfnamefont{C.~A.} \bibnamefont{Bertulani}},
  \bibinfo{journal}{Phys. Rev. Lett.} \textbf{\bibinfo{volume}{94}},
  \bibinfo{pages}{072701} (\bibinfo{year}{2005}).

\bibitem[{\citenamefont{Baye et~al.}(2005)\citenamefont{Baye, Capel, and
  Goldstein}}]{BCG05}
\bibinfo{author}{\bibfnamefont{D.}~\bibnamefont{Baye}},
  \bibinfo{author}{\bibfnamefont{P.}~\bibnamefont{Capel}}, \bibnamefont{and}
  \bibinfo{author}{\bibfnamefont{G.}~\bibnamefont{Goldstein}},
  \bibinfo{journal}{Phys. Rev. Lett.} \textbf{\bibinfo{volume}{95}},
  \bibinfo{pages}{082502} (\bibinfo{year}{2005}).

\bibitem[{\citenamefont{Ogata and Bertulani}(2010)}]{OB10}
\bibinfo{author}{\bibfnamefont{K.}~\bibnamefont{Ogata}} \bibnamefont{and}
  \bibinfo{author}{\bibfnamefont{C.~A.} \bibnamefont{Bertulani}},
  \bibinfo{journal}{Prog. of Theor. Phys.} \textbf{\bibinfo{volume}{123}},
  \bibinfo{pages}{701} (\bibinfo{year}{2010}).

\bibitem[{\citenamefont{Fukui et~al.}(2014)\citenamefont{Fukui, Ogata, and
  Capel}}]{FOC14}
\bibinfo{author}{\bibfnamefont{T.}~\bibnamefont{Fukui}},
  \bibinfo{author}{\bibfnamefont{K.}~\bibnamefont{Ogata}}, \bibnamefont{and}
  \bibinfo{author}{\bibfnamefont{P.}~\bibnamefont{Capel}},
  \bibinfo{journal}{Phys.~Rev.~C} \textbf{\bibinfo{volume}{90}},
  \bibinfo{pages}{034617} (\bibinfo{year}{2014}).

\bibitem[{\citenamefont{Aguiar et~al.}(1997)\citenamefont{Aguiar, Zardi, and
  Vitturi}}]{AZV97}
\bibinfo{author}{\bibfnamefont{C.~E.} \bibnamefont{Aguiar}},
  \bibinfo{author}{\bibfnamefont{F.}~\bibnamefont{Zardi}}, \bibnamefont{and}
  \bibinfo{author}{\bibfnamefont{A.}~\bibnamefont{Vitturi}},
  \bibinfo{journal}{Phys. Rev. C} \textbf{\bibinfo{volume}{56}},
  \bibinfo{pages}{1511} (\bibinfo{year}{1997}).

\bibitem[{\citenamefont{Hebborn and Capel}(2017{\natexlab{a}})}]{HC17}
\bibinfo{author}{\bibfnamefont{C.}~\bibnamefont{Hebborn}} \bibnamefont{and}
  \bibinfo{author}{\bibfnamefont{P.}~\bibnamefont{Capel}},
  \bibinfo{journal}{Phys. Rev. C} \textbf{\bibinfo{volume}{96}},
  \bibinfo{pages}{054607} (\bibinfo{year}{2017}{\natexlab{a}}).

\bibitem[{\citenamefont{Wallace}(1973)}]{Wal73}
\bibinfo{author}{\bibfnamefont{S.~J.} \bibnamefont{Wallace}},
  \bibinfo{journal}{Phys.~Rev.~D} \textbf{\bibinfo{volume}{8}},
  \bibinfo{pages}{1846} (\bibinfo{year}{1973}).

\bibitem[{\citenamefont{Brooke}(1999)}]{BrookePhD}
\bibinfo{author}{\bibfnamefont{J.~M.} \bibnamefont{Brooke}}, Ph.D. thesis,
  \bibinfo{school}{University of Surrey}, \bibinfo{address}{Surrey}
  (\bibinfo{year}{1999}).

\bibitem[{\citenamefont{Brooke et~al.}(1999)\citenamefont{Brooke, Al-Khalili,
  and Tostevin}}]{BAT99}
\bibinfo{author}{\bibfnamefont{J.~M.} \bibnamefont{Brooke}},
  \bibinfo{author}{\bibfnamefont{J.~S.} \bibnamefont{Al-Khalili}},
  \bibnamefont{and} \bibinfo{author}{\bibfnamefont{J.~A.}
  \bibnamefont{Tostevin}}, \bibinfo{journal}{Phys. Rev. C}
  \textbf{\bibinfo{volume}{59}}, \bibinfo{pages}{1560} (\bibinfo{year}{1999}).

\bibitem[{\citenamefont{Margueron et~al.}(2003)\citenamefont{Margueron,
  Bonaccorso, and Brink}}]{MBB03}
\bibinfo{author}{\bibfnamefont{J.}~\bibnamefont{Margueron}},
  \bibinfo{author}{\bibfnamefont{A.}~\bibnamefont{Bonaccorso}},
  \bibnamefont{and} \bibinfo{author}{\bibfnamefont{D.}~\bibnamefont{Brink}},
  \bibinfo{journal}{Nucl. Phys. A} \textbf{\bibinfo{volume}{720}},
  \bibinfo{pages}{337 } (\bibinfo{year}{2003}).

\bibitem[{\citenamefont{Capel et~al.}(2008)\citenamefont{Capel, Baye, and
  Suzuki}}]{CBS08}
\bibinfo{author}{\bibfnamefont{P.}~\bibnamefont{Capel}},
  \bibinfo{author}{\bibfnamefont{D.}~\bibnamefont{Baye}}, \bibnamefont{and}
  \bibinfo{author}{\bibfnamefont{Y.}~\bibnamefont{Suzuki}},
  \bibinfo{journal}{Phys. Rev. C} \textbf{\bibinfo{volume}{78}},
  \bibinfo{pages}{054602} (\bibinfo{year}{2008}).

\bibitem[{\citenamefont{Lenzi et~al.}(1995)\citenamefont{Lenzi, Vitturi, and
  Zardi}}]{LVZ95}
\bibinfo{author}{\bibfnamefont{S.~M.} \bibnamefont{Lenzi}},
  \bibinfo{author}{\bibfnamefont{A.}~\bibnamefont{Vitturi}}, \bibnamefont{and}
  \bibinfo{author}{\bibfnamefont{F.}~\bibnamefont{Zardi}}, \bibinfo{journal}{Z.
  Phys. A} \textbf{\bibinfo{volume}{352}}, \bibinfo{pages}{303}
  (\bibinfo{year}{1995}).

\bibitem[{\citenamefont{Hebborn and Capel}(2017{\natexlab{b}})}]{HCp17}
\bibinfo{author}{\bibfnamefont{C.}~\bibnamefont{Hebborn}} \bibnamefont{and}
  \bibinfo{author}{\bibfnamefont{P.}~\bibnamefont{Capel}}, in
  \emph{\bibinfo{booktitle}{Proc. of the 55th International Winter Meeting on
  Nuclear Physics}}, edited by
  \bibinfo{editor}{\bibfnamefont{C.}~\bibnamefont{Sfienti}},
  \bibinfo{editor}{\bibfnamefont{L.}~\bibnamefont{Fabbietti}},
  \bibnamefont{and} \bibinfo{editor}{\bibfnamefont{W.}~\bibnamefont{K\"{u}hn}}
  (\bibinfo{publisher}{PoS}, \bibinfo{address}{Bormio, Italy},
  \bibinfo{year}{2017}{\natexlab{b}}).

\bibitem[{\citenamefont{Brink}(1985)}]{B85}
\bibinfo{author}{\bibfnamefont{D.~M.} \bibnamefont{Brink}},
  \emph{\bibinfo{title}{Semi-classical methods in nucleus-nucleus scattering}}
  (\bibinfo{publisher}{Cambridge University Press, Cambridge},
  \bibinfo{year}{1985}).

\bibitem[{\citenamefont{Summers et~al.}(2002)\citenamefont{Summers, Al-Khalili,
  and Johnson}}]{SAJ02}
\bibinfo{author}{\bibfnamefont{N.~C.} \bibnamefont{Summers}},
  \bibinfo{author}{\bibfnamefont{J.~S.} \bibnamefont{Al-Khalili}},
  \bibnamefont{and} \bibinfo{author}{\bibfnamefont{R.~C.}
  \bibnamefont{Johnson}}, \bibinfo{journal}{Phys. Rev. C}
  \textbf{\bibinfo{volume}{66}}, \bibinfo{pages}{014614}
  (\bibinfo{year}{2002}).

\bibitem[{\citenamefont{Capel et~al.}(2004)\citenamefont{Capel, Goldstein, and
  Baye}}]{CGB04}
\bibinfo{author}{\bibfnamefont{P.}~\bibnamefont{Capel}},
  \bibinfo{author}{\bibfnamefont{G.}~\bibnamefont{Goldstein}},
  \bibnamefont{and} \bibinfo{author}{\bibfnamefont{D.}~\bibnamefont{Baye}},
  \bibinfo{journal}{Phys. Rev. C} \textbf{\bibinfo{volume}{70}},
  \bibinfo{pages}{064605} (\bibinfo{year}{2004}).

\bibitem[{\citenamefont{Thompson}(1988)}]{FRESCO}
\bibinfo{author}{\bibfnamefont{I.~J.} \bibnamefont{Thompson}},
  \bibinfo{journal}{Comput. Phys. Rep.} \textbf{\bibinfo{volume}{7}},
  \bibinfo{pages}{167} (\bibinfo{year}{1988}).

\bibitem[{\citenamefont{Koning and Delaroche}(2003)}]{KD03}
\bibinfo{author}{\bibfnamefont{A.}~\bibnamefont{Koning}} \bibnamefont{and}
  \bibinfo{author}{\bibfnamefont{J.}~\bibnamefont{Delaroche}},
  \bibinfo{journal}{Nucl. Phys. A.} \textbf{\bibinfo{volume}{713}},
  \bibinfo{pages}{231 } (\bibinfo{year}{2003}).

\end{thebibliography}

\end{document}